\documentclass[a4paper,USenglish,cleveref,autoref,thm-restate]{lipics-v2021}

\title{A 2.05-Approximation for Unit Disk Graph Coloring}
\titlerunning{A 2.05-Approximation for Unit Disk Graph Coloring}

\author{Guilherme D. da Fonseca}{Aix-Marseille Universit\'e and LIS, France \and \url{https://pageperso.lis-lab.fr/guilherme.fonseca/}}{guilherme.fonseca@lis-lab.fr}{https://orcid.org/0000-0002-9807-028X}{}
\author{Kunal Dutta}{Institute of Informatics, University of Warsaw, Poland \and \url{https://www.mimuw.edu.pl/~kdutta/index.html}}{kdutta@mimuw.edu.pl}{https://orcid.org/0000-0003-3055-9326}{Supported by the Polish NCN Opus grant nr. 2023/51/B/ST6/02989}
\authorrunning{G.\,D. da Fonseca and K. Dutta}
\Copyright{Guilherme D. da Fonseca and Kunal Dutta}

\ccsdesc[500]{Theory of computation~Computational geometry}
\ccsdesc[500]{Theory of computation~Rounding techniques}
\ccsdesc[300]{Mathematics of computing~Graph coloring}

\keywords{unit disk graphs, unit square graphs, graph coloring, approximation algorithms,
fractional chromatic number, randomized rounding, Lov\'asz Local Lemma}

\EventEditors{}
\EventNoEds{0}
\EventLongTitle{}
\EventShortTitle{}
\EventAcronym{}
\EventYear{}
\EventDate{}
\EventLocation{}
\EventLogo{}
\SeriesVolume{}
\ArticleNo{}
\hideLIPIcs

\newcommand{\chif}{\chi_f}
\newcommand{\inorm}[1]{\lVert #1\rVert_\infty}
\newcommand{\R}{\mathbb{R}}

\newcommand{\Z}{\mathbb{Z}}
\newcommand{\Sref}[1]{\hyperref[cond:S#1]{(S#1)}}
\newcommand{\cI}{\mathcal{I}}
\newcommand{\Exp}{\mathbb{E}}

\begin{document}
\nolinenumbers

\maketitle

\begin{abstract}
The best known approximation ratio for coloring unit disk graphs has been $3$ since the early 1990s.
It comes from a left-to-right sweep, in which the earlier neighbors of a vertex are covered by three cliques.
We give a randomized polynomial-time algorithm that, given the points, colors every unit disk graph $G$ with at most $2.05\,\chi(G)$ colors.
The algorithm first approximately solves the LP for the fractional chromatic number $\chif$.
If $\chif$ is large, it samples color classes from the LP solution independently in each cell of a coarse grid and colors the remaining vertices with the sweep, reusing a sampled color whenever no neighbor has it.
This refines the Round-and-Approx scheme of Bansal, Caprara, and Sviridenko; the new ingredient is locality: a bounded-differences inequality combined with the constructive Lov\'asz Local Lemma controls all residual cliques simultaneously.
If $\chif$ is small, then so is the clique number, and dynamic programming provides good approximations.
Our techniques also improve the best known approximation ratios for intersection graphs of translates of a convex polygon, from $3$ to $2.05$, and for unit square graphs, from $2$ to $1.68$.
\end{abstract}

\section{Introduction}

A \emph{unit disk graph} has a vertex for each of the $n$ points of a finite set $P \subset \R^2$ and an edge between two points at Euclidean distance at most $1$; equivalently, it is the intersection graph of closed disks of diameter $1$.
A \emph{unit square graph} is defined in the same way with the $L_\infty$ distance, and is the intersection graph of closed axis-parallel unit squares.

We recall the few graph notions we need.
A \emph{(proper) coloring} assigns a color to every vertex so that adjacent vertices receive distinct colors, that is, intersecting disks or squares get distinct colors; the \emph{chromatic number} $\chi(G)$ is the minimum number of colors.
A \emph{clique} is a set of pairwise adjacent vertices (pairwise intersecting objects), and the \emph{clique number} $\omega(G)$ is the size of a largest clique.
An \emph{independent set} is a set of pairwise non-adjacent vertices (pairwise disjoint objects), and $\alpha(G)$ is the size of a largest one.
The vertices of a clique need distinct colors and every color class is an independent set, so $\chi(G) \ge \omega(G)$ and $\chi(G) \ge n/\alpha(G)$.
An algorithm is a \emph{$\rho$-approximation} if it always uses at most $\rho\,\chi(G)$ colors.
We use the real RAM model of computation.

Coloring these graphs models frequency assignment and has been studied at least since 1990~\cite{ClarkCJ90}.
Deciding $k$-colorability is NP-complete for unit disk graphs for every fixed $k \ge 3$~\cite{ClarkCJ90,GrafSW98}, and likewise for unit square graphs~\cite{CaoduroS24}.
Hence neither class admits a $(4/3 - \varepsilon)$-approximation unless $\mathrm{P} = \mathrm{NP}$.

\subparagraph*{The factors 3 and 2.}
Sort the points by $x$-coordinate and color greedily (first-fit) in that order.
When a vertex $v$ is colored, its already colored neighbors lie in the left half of the unit disk around $v$.
This half-disk is the union of three closed $60^\circ$ sectors, each of diameter $1$, so each sector together with $v$ induces a clique.
Hence at most $3(\omega-1)$ earlier neighbors exist, and the sweep uses at most $3\omega - 2$ colors.
Since $\chi \ge \omega$, this is a $3$-approximation, which goes back to Peeters~\cite{Peeters91} (see also~\cite{MaratheBHRR95,GrafSW98}).
For unit squares the left half of the neighborhood is a $1 \times 2$ rectangle, the union of two unit squares, which gives the bound $\chi \le 2\omega - 1$ of Perepelitsa~\cite{Perepelitsa03} and a $2$-approximation.

Both factors can also be obtained by strip decompositions.
The STRIPE algorithm of Gr\"af, Stumpf, and Wei\ss{}enfels~\cite{GrafSW98} partitions the points into vertical stripes of width at most $\sqrt3/2$.
The disks in one stripe form a \emph{cocomparability graph}, the complement of a graph with a transitive orientation, and such graphs can be colored optimally in polynomial time; the stripe colorings are then combined by a matching technique, with performance ratio~$3$.
The squares centered in a horizontal strip of height~$1$ intersect exactly when their $x$-coordinates differ by at most $1$, so they form a \emph{unit interval graph}, the intersection graph of unit intervals on a line.
Such graphs are \emph{perfect}: $\chi = \omega$ holds for them and for all their induced subgraphs, and they can be colored optimally by sorting.
Strips two apart contain no adjacent pairs, so alternating palettes on strips of odd and even index use at most $2\chi$ colors, as in a sum-coloring algorithm~\cite{BorodinIYZ12}.
Neither approach goes below the factor of the sweep.
We are not aware of any improvement of the ratio $3$ since 1991, and improving the constant in $\chi \le 3\omega - 2$ is a longstanding open problem~\cite{GrafSW98,EsperetJdM23}.
The literature on unit square graphs is thinner~\cite{DumitrescuJ12,CaoduroS24}, and we are not aware of any algorithm with a ratio below $2$.

Any analysis that compares the number of colors with $\omega$ runs into the open question of how large $\chi/\omega$ can be.
For unit disk graphs it is known only that $\limsup_{\omega \to \infty} \chi/\omega \in [3/2, 3]$~\cite{MalesinskaPW98,Peeters91}.
Moreover, the \emph{degeneracy} (the largest $d$ such that some subgraph has all degrees at least $d$; greedy coloring in a suitable order uses at most $d + 1$ colors) can be as large as $3\omega - 3$~\cite{KimKN04}.
For unit square graphs $\chi/\omega < 2$ always~\cite{Perepelitsa03}, while the published lower bound for large $\omega$ is $5/4$~\cite{DumitrescuJ12,CaoduroS24}, which we raise to $4/3$ in \cref{sec:squares}.
An algorithm analyzed against $\omega$ that beats the factor $3$ would improve the constant in $\chi \le 3\omega - 2$, so we compare instead with a lower bound on $\chi$ that is stronger than $\omega$ and computable in polynomial time: the fractional chromatic number.

The \emph{fractional chromatic number} $\chif(G)$ is the optimum of the linear-programming relaxation of coloring: put a nonnegative weight on every independent set so that the sets containing each vertex have total weight at least $1$, and minimize the total weight.
A coloring is such a weighting with weights $0$ and $1$, and a clique meets each independent set at most once, so $\omega \le \chif \le \chi$, and $\chif \le 2.155\,\omega$ for unit disk graphs~\cite{GerkeM01}.
The weighted version of $\chif$ is NP-hard on unit disk graphs, since by the ellipsoid method it is as hard as maximum-weight independent set~\cite{GrotschelLS88,Jansen03}.
A $2$-approximation for $\chif$ of unit disk graphs is known~\cite{Matsui00}, and Jansen~\cite{Jansen03} showed that the weighted version has a PTAS whenever maximum-weight independent set does; Erlebach, Jansen, and Seidel~\cite{ErlebachJS05} gave such a PTAS for disks and for squares.
(\Cref{lem:lp} gives a short direct algorithm for the unweighted case.)

\subparagraph*{Our results.}
Our main result improves the factor $3$ for disks.

\begin{theorem}\label{thm:main}
There is a randomized algorithm with expected polynomial running time that, given a unit disk graph $G$ by its points, outputs a proper coloring of $G$ with at most $2.0483\,\chi(G)$ colors.
\end{theorem}

We also prove a bound in terms of $\chif$, at the cost of an additive constant.

\begin{theorem}\label{thm:frac}
There is a constant $K$ and a randomized algorithm with expected polynomial running time that, given a unit disk graph $G$ by its points, outputs a proper coloring of $G$ with at most $2.0483\,\chif(G) + K$ colors.
\end{theorem}

The proof has three parts: rounding an approximately optimal fractional coloring, reusing its colors, and coloring graphs of bounded clique number.
The first part draws color classes of the fractional coloring and colors the vertices they do not cover with the sweep, using new colors.
This gives $(1 + \ln 3 + \varepsilon)\chif + K_\varepsilon$ colors for every $\varepsilon \in (0,1]$ (\cref{lem:general}).
The second part lets a remaining vertex take, when possible, a color of the rounding that none of its neighbors received.
The number of such colors is controlled by the same Local Lemma argument, using that the neighborhood of a vertex contains at most $5$ pairwise non-adjacent vertices.
This lowers the constant from $1 + \ln 3 > 2.0986$ to $2.04822$ and proves \cref{thm:frac} (\cref{sec:reuse}).
The third part removes the additive constant $K$, which is a small fraction of $\chif$ once $\chif$ is large, so it matters only when $\chif$, and hence $\omega \le \chif$, is below a constant.
Then, within a vertical slab of width $1$ sorted by $y$, each point is adjacent only to points a bounded number of positions away, so dynamic programming colors each slab optimally, and two alternating palettes give at most $2\chi$ colors (\cref{lem:slabs}).
Since $2 < 2.0483$, running the appropriate algorithm in each case proves \cref{thm:main} (\cref{sec:reuse}).
Without reuse, the same argument colors every unit disk graph with at most $(1 + \ln 3 + \varepsilon)\chi$ colors (\cref{lem:absolute}), which is already a $2.1$-approximation since $1 + \ln 3 < 2.0987$.
The thresholds are large: with our bounds, the rounding beats the sweep for disks only for $\chif$ above about $100$ and reuse beats $1 + \ln 3$ only above about $10^7$, and below such thresholds the guarantee comes from \cref{lem:slabs}, whose running time is exponential in the threshold.
With the constants of \cref{thm:main}, the LP of \cref{lem:lp} has up to about $n^{10^{16}}$ variables and \cref{lem:slabs} is used for $\chif$ up to about $3 \cdot 10^{18}$; we did not try to improve the degree of the polynomial.

\subparagraph*{Other classes.}
The rounding and the slabs work for every \emph{$c$-sweepable} class (\cref{def:sweepable,lem:general,lem:absolute,thm:reuse}), which captures exactly the geometry used by the sweep: without reuse the constant is $1 + \ln c$ in place of $1 + \ln 3$, and the bound $(1 + \ln c + \varepsilon)\chi$ holds if $c \ge 3$.
Unit disk graphs are $3$-sweepable.
Intersection graphs of translates of a convex polygon are $3$-sweepable as well, with at most $5$ pairwise non-adjacent neighbors per vertex (\cref{sec:translates,sec:reuse}), which gives \cref{thm:translates}.
For these graphs, Kim, Kostochka, and Nakprasit~\cite{KimKN04} proved $\chi \le 3\omega - 2$ by a sweep.

\begin{theorem}\label{thm:translates}
There is a randomized algorithm with expected polynomial running time that, given a convex polygon and the translation vectors of the intersection graph $G$ of its translates, outputs a proper coloring of $G$ with at most $2.0483\,\chi(G)$ colors.
There is also such an algorithm that uses at most $2.0483\,\chif(G) + K$ colors, where $K$ is a constant.
\end{theorem}

Unit square graphs are $2$-sweepable, but for them the bound $2\chi$ of the slabs is no better than the sweep.
\Cref{sec:squares} lets consecutive slabs overlap and turns the coloring of one slab into that of the next, starting in the overlap and proceeding one strip of width $1$ at a time, which uses at most $\lceil 3\chi/2 \rceil$ colors when $\omega$ is bounded.

\begin{theorem}\label{thm:squares}
There is a randomized algorithm with expected polynomial running time that, given a unit square graph $G$ by its points, outputs a proper coloring of $G$ with at most $1.6796\,\chi(G)$ colors.
There is also such an algorithm that uses at most $1.6796\,\chif(G) + K$ colors, where $K$ is a constant.
\end{theorem}

\begin{figure}[t]
\centering
\includegraphics{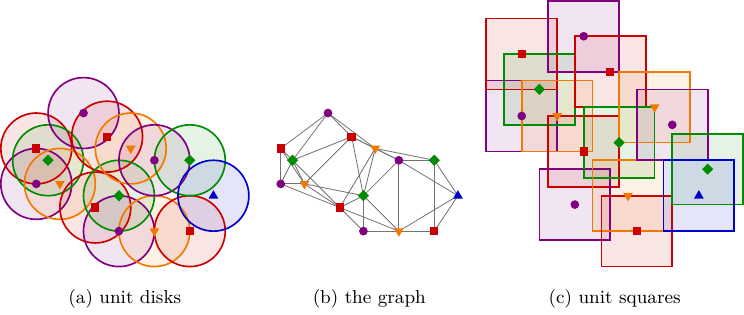}
\caption{A graph with $\omega = \chif = 4$ and $\chi = 5$: (a) disks of diameter $1$, (b) the graph on the disk centers, (c) unit squares; colors show a $5$-coloring, and some adjacent disks and squares only touch.
Numbering the vertices $0, \dots, 14$ as below, the sets $\{2,9,10\}$, $\{0,4,8,14\}$, $\{1,5,6,12\}$, $\{2,8,13\}$, $\{3,4,7,11\}$, $\{1,5,11,12\}$, $\{0,6,10,13\}$, $\{3,7,9,14\}$ are independent and cover each vertex twice.
Disk centers $\times 6$: $(0,7)$, $(0,10)$, $(1,9)$, $(2,7)$, $(4,13)$, $(5,5)$, $(6,11)$, $(7,3)$, $(7,6)$, $(8,10)$, $(10,3)$, $(10,9)$, $(13,3)$, $(13,9)$, $(15,6)$.
Square centers $\times 8$: $(0,13)$, $(0,20)$, $(2,16)$, $(4,13)$, $(7,22)$, $(7,9)$, $(10,18)$, $(6,3)$, $(11,10)$, $(15,14)$, $(12,4)$, $(17,12)$, $(13,0)$, $(21,7)$, $(20,4)$.}
\label{fig:gap}
\end{figure}

\subparagraph*{Fractional versus integral chromatic number.}
\Cref{thm:frac,thm:translates,thm:squares} show that $\chi/\chif$ is asymptotically at most $2.0483$ for unit disk graphs and for translates of a convex polygon, and $1.6796$ for unit square graphs.
We do not know whether $\chi/\chif$ is asymptotically larger than $1$.
For small $\chi$ it is: every odd cycle $C_{2k+1}$ is both a unit disk graph and a unit square graph, with $\chi(C_{2k+1}) = 3$ and $\chif(C_{2k+1}) = 2 + 1/k$.
The ratio $3/(2 + 1/k)$ is $6/5$ for $C_5$ and tends to $3/2$ as $k \to \infty$.
So a bound $\chi \le \rho\,\chif$ with $\rho < 3/2$ needs an additive constant.
The graph in \cref{fig:gap}, which is both a unit disk graph and a unit square graph, shows that the additive gap $\chi - \chif$ can reach $1$.
We do not know whether $\chi - \chif$ can exceed $1$ in these classes; such a graph would need $\chi \ge \omega + 2$.
If $\chi - \chif = o(\chif)$ in these classes, then $\chi/\chif$ tends to $1$, and the factors of \cref{thm:frac,thm:squares} are a limitation of the rounding and not of the fractional coloring.
It is then natural to ask for an algorithm that uses $(1 + \varepsilon)\chif + K_\varepsilon$ colors.

\subparagraph*{Technique and related work.}
Sampling from an LP solution and finishing the rest with a $c$-approximation is the Round-and-Approx framework of Bansal, Caprara, and Sviridenko~\cite{BansalCS09}, who applied it to two-dimensional vector bin packing.
There a union bound over a few global quantities suffices.
For coloring, all $n$ local cliques must be controlled at once, and a union bound over them requires $\chi = \Omega(\log n)$, the regime of an algorithm for circular-arc coloring~\cite{Kumar98}.
We avoid this by sampling \emph{independently in each cell of a grid}, so that the Lov\'asz Local Lemma~\cite{ErdosL75}, in the algorithmic form of Moser and Tardos~\cite{MoserT10}, applies with a dependency degree independent of $n$.
Rounding controlled by the Local Lemma is classical for column-sparse integer programs~\cite{LeightonLRS01,Srinivasan06} and for edge coloring of multigraphs~\cite{Kahn96,MolloyR02}; here the geometry supplies the sparsity.
Reusing colors of the rounding for the remaining vertices lowers the constants, because the neighborhood of a vertex contains few pairwise non-adjacent vertices.
Graphs with small $\chif$ are colored with slabs, as in the strip decompositions of~\cite{GrafSW98,BorodinIYZ12}, except that bounded clique number lets dynamic programming color each slab optimally; for unit squares, overlapping slabs replace the two alternating palettes.

Fractional colorings of random geometric graphs have been rounded locally~\cite{McDiarmidM11}, by solving the coloring LP on blocks of a grid much as the proof of \cref{lem:lp} does, but that bound compares $\chi$ with $\chif$ of a graph with a larger distance threshold.
Localized random independent sets also motivate a derandomized argument for distributed coloring of unit disk graphs~\cite{EsperetJdM23}.

\subparagraph*{Organization.}
\Cref{sec:prelim} defines $c$-sweepable classes, shows that unit disk and unit square graphs are $3$- and $2$-sweepable, and collects the tools we use: the sweep, shifted tilings, the fractional coloring LP, bounded differences, and the Moser--Tardos algorithm.
\Cref{sec:rounding} rounds the fractional coloring and proves the bound $(1 + \ln c + \varepsilon)\chif + K_\varepsilon$ (\cref{lem:general}).
\Cref{sec:absolute} handles bounded clique number and gives the $2.1$-approximation (\cref{lem:absolute}) for unit disk graphs and for translates of a convex polygon, which \cref{sec:translates} shows to be $3$-sweepable.
\Cref{sec:reuse} reuses colors, improves the leading constants, and proves \cref{thm:main,thm:frac}.
\Cref{sec:squares} treats unit square graphs: it colors them with overlapping slabs when $\omega$ is bounded (\cref{lem:stitch}), proves \cref{thm:squares}, and gives the lower bound $4/3$ on $\limsup \chi/\omega$.

\section{Preliminaries}\label{sec:prelim}

All graphs are finite and simple.
Points of $P$ may coincide, so $P$ is a multiset, and ties in the lexicographic order are broken arbitrarily.
For a graph $G$, $\omega(G)$, $\chi(G)$, and $\cI(G)$ denote the clique number, the chromatic number, and the family of independent sets.
A \emph{fractional coloring} of $G$ is a vector $y \in \R_{\ge 0}^{\cI(G)}$ with $\sum_{I \ni v} y_I \ge 1$ for every vertex $v$.
Its \emph{value} is $\sum_I y_I$, and $\chif(G)$ is the minimum value.
Every clique meets every independent set in at most one vertex, so $\omega \le \chif \le \chi$.
For $A \subseteq \R^2$ we write $\operatorname{diam}_\infty(A) = \sup_{a,b \in A} \inorm{a-b}$.

\begin{definition}[$c$-sweepable]\label{def:sweepable}
Let $c \ge 2$ be an integer.
A class $\mathcal G$ of graphs whose vertices are points of $\R^2$ is \emph{$c$-sweepable} if every $G = (P,E) \in \mathcal G$ satisfies:
\begin{enumerate}[(S1)]
\item\label{cond:S1} if $uv \in E$ then $\inorm{u-v} \le 1$;
\item\label{cond:S2} every $A \subseteq P$ with $\operatorname{diam}_\infty(A) \le 1/2$ is a clique of $G$;
\item\label{cond:S3} there is a linear order $\prec$ on $P$ and, for every $v \in P$, sets $Q_1(v), \dots, Q_c(v) \subseteq \R^2$ such that $v \in Q_j(v)$, $\operatorname{diam}_\infty(Q_j(v)) \le 1$, $P \cap Q_j(v)$ is a clique of $G$, and every neighbor $u \prec v$ of $v$ lies in $\bigcup_j Q_j(v)$.
  The order and membership in $Q_j(v)$ can be computed in polynomial time.
\end{enumerate}
\end{definition}

The conditions are inherited by induced subgraphs: for $R \subseteq P$, the graph $G[R]$ satisfies \Sref{1}--\Sref{3} with the restricted order and the same sets $Q_j(v)$.
Condition \Sref{3} alone implies that $G$ is \emph{$c$-simplicial}~\cite{KammerT14,YeB12}, which already bounds first-fit in the order $\prec$ by $c(\omega-1)+1$ colors; \Sref{1}, \Sref{2}, and the sets $Q_j(v)$ add the locality that the rounding needs.

\begin{lemma}\label{lem:classes}
Unit disk graphs are $3$-sweepable and unit square graphs are $2$-sweepable.
\end{lemma}

\begin{figure}[t]
\centering
\includegraphics{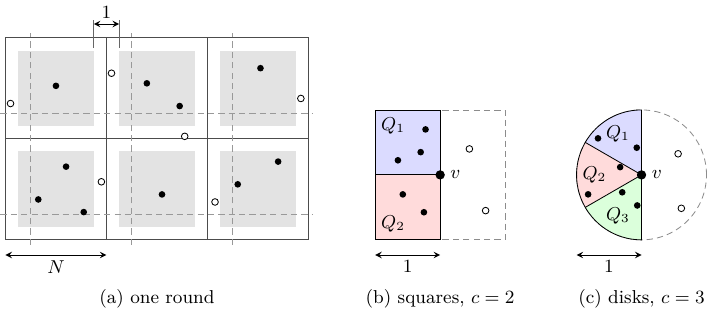}
\caption{(a) The tiles of one round (solid) and of the next (dashed); points in the band around the interiors (shaded) are hollow.
(b), (c) The sets $Q_j(v)$ cover the earlier neighbors of $v$ (filled); later ones (hollow) lie in the right half.}
\label{fig:tiles}
\end{figure}

\begin{proof}
Let $\prec$ be the lexicographic order on $(x,y)$.
\Sref{1}: in both classes adjacent points are at Euclidean or $L_\infty$ distance at most $1$, so their $L_\infty$ distance is at most $1$.
\Sref{2}: two points of a set of $L_\infty$-diameter at most $1/2$ are at Euclidean distance at most $\sqrt 2/2 < 1$, so they are adjacent in both classes.

\Sref{3} for squares: every neighbor $u \prec v$ has $x_u \le x_v$ and $|y_u - y_v| \le 1$, so it lies in $Q_1(v) = [x_v - 1, x_v] \times [y_v, y_v + 1]$ or $Q_2(v) = [x_v-1,x_v] \times [y_v - 1, y_v]$.
Both are closed unit squares containing $v$ (\cref{fig:tiles}(b)), and points in a closed unit square are pairwise at $L_\infty$ distance at most $1$, so they form a clique.

\Sref{3} for disks: every neighbor $u \prec v$ lies in the closed left half-disk $\{p : \|p - v\|_2 \le 1,\ x_p \le x_v\}$.
If $x_u = x_v$ then $y_u \le y_v$, so $u$ lies on the downward ray.
Let $Q_1(v), Q_2(v), Q_3(v)$ be the closed sectors of radius $1$ with apex $v$ spanning the directions $[90^\circ, 150^\circ]$, $[150^\circ, 210^\circ]$, and $[210^\circ, 270^\circ]$.
They cover the closed left half-disk, including both boundary rays (\cref{fig:tiles}(c)).
A closed $60^\circ$ sector of radius $1$ has Euclidean diameter exactly $1$, so its points form a clique, and its $L_\infty$-diameter is at most its Euclidean diameter.
Membership is decided without square roots, since for $u - v = (a,b)$ with $a \le 0$ the rays at $150^\circ$ and $210^\circ$ are where $a^2 = 3b^2$.
\end{proof}

\begin{lemma}[the sweep]\label{lem:sweep}
Let $G = (P,E)$ belong to a $c$-sweepable class and let $R \subseteq P$.
First-fit on $G[R]$ in the order $\prec$ is a proper coloring with at most
\[
  1 + \max_{v \in R} \sum_{j=1}^c |(Q_j(v) \cap R) \setminus \{v\}|
  \;\le\; c\,(\omega(G[R]) - 1) + 1
\]
colors.
In particular $\chi(G) \le c\,\omega(G) - c + 1$.
\end{lemma}

\begin{proof}
When $v$ is colored, the colors on its earlier neighbors in $R$ number at most the number of such neighbors, and these lie in $\bigcup_j (Q_j(v) \cap R) \setminus \{v\}$ by \Sref{3}.
Each $Q_j(v) \cap R$ is a clique containing $v$, so it has at most $\omega(G[R]) - 1$ other points.
\end{proof}

\subparagraph*{Tilings.}
For integers $L \ge 2$ and $o$, the \emph{tiling} $\mathcal T(L, o)$ consists of the tiles
\[
  T = [o + aL,\ o + (a+1)L) \times [o + bL,\ o + (b+1)L), \qquad a, b \in \Z,
\]
with \emph{interiors} $T^\circ = [o + aL + \tfrac12,\ o + (a+1)L - \tfrac12) \times [o + bL + \tfrac12,\ o + (b+1)L - \tfrac12)$ (\cref{fig:tiles}(a)).
The offset $o$ shifts the tiling along the diagonal, in the spirit of the shifting technique~\cite{HochbaumM85}.

\begin{lemma}\label{lem:tiles}
Let $L \ge 2$ be an integer.
\begin{enumerate}[(i)]
\item\label{lem:tiles:i} Two points in the interiors of different tiles of $\mathcal T(L, o)$ are at $L_\infty$ distance more than $1$.
\item\label{lem:tiles:ii} Among any $L$ consecutive integers $o$, there are at most two for which a given point $p$ lies in no interior of $\mathcal T(L, o)$.
\end{enumerate}
\end{lemma}

\begin{proof}
(i) The two points differ in one coordinate by more than $(o + a'L + \tfrac12) - (o + a'L - \tfrac12) = 1$ for some integer $a'$, because the interiors are half-open.
(ii) The point $p$ lies in no interior only if $(x_p - o) \bmod L$ or $(y_p - o) \bmod L$ lies in $[L-\tfrac12, L) \cup [0, \tfrac12)$, a half-open arc of length $1$ on a circle of length $L$.
As $o$ ranges over $L$ consecutive integers, the values $(x_p - o) \bmod L$ are spaced $1$ apart, so exactly one of them lies in the arc.
The same holds for $y_p$.
\end{proof}

\subparagraph*{Model of computation.}
We use the real RAM without the floor function.
The input coordinates enter only through adjacency, the order $\prec$, membership in the sets $Q_j(v)$, and the placement of points in tiles and slabs; everything else, including the linear programs, is exact arithmetic on rationals of polynomial bit size, so for rational input the algorithms also run in polynomial time on a Turing machine.
All our algorithms color each connected component of $G$ separately, with the same colors, which suffices because $\omega$, $\chif$, and $\chi$ of $G$ are the maxima over its components.
Translating all points and all sets $Q_j(v)$ by a common vector preserves \Sref{1}--\Sref{3}, so we may translate a component to have least $x$- and $y$-coordinates $0$.
By \Sref{1} its coordinates then lie in $[0, n-1]$, and the integer part of each is found by binary search over $\{0, \dots, n-1\}$, which places every point in its tile or slab with $O(\log n)$ comparisons.

\begin{lemma}[fractional coloring]\label{lem:lp}
For every $c$-sweepable class and every fixed $\delta > 0$ there is a polynomial-time algorithm that, given a graph $G$ of the class, computes a fractional coloring $y$ of value at most $(1+\delta)\chif(G)$ with rational entries and polynomially many non-zero entries.
\end{lemma}

\begin{proof}
Let $D = 2 + \lceil 2/\delta \rceil$, so that $D/(D-2) \le 1 + \delta$.
Fix an offset $o \in \{0, \dots, D-1\}$ and a tile $T$ of $\mathcal T(D, o)$ with $P_T := P \cap T^\circ \neq \emptyset$.
The interior $T^\circ$ is covered by $(2D-2)^2$ half-open cells of side $1/2$, and by \Sref{2} each cell contains at most one vertex of an independent set.
Hence $G[P_T]$ has at most $(n+1)^{4D^2}$ independent sets, and its fractional coloring LP has polynomial size.
Its constraint matrix has entries $0$ and $1$ and its right-hand side is all ones, whatever the coordinates, so an optimal basic solution $y^T$ can be computed exactly in polynomial time.
It is rational, has at most $|P_T|$ non-zero entries, and has value $f_T = \chif(G[P_T]) \le \chif(G)$.

Next we merge the solutions $y^T$ over the tiles of $\mathcal T(D, o)$.
For each $T$, split $[0, f_T)$ into consecutive half-open intervals, one of length $y^T_I$ for each set $I$ with $y^T_I > 0$.
For $x \ge 0$ let $S(x)$ be the union, over all tiles $T$ with $x < f_T$, of the set $I$ whose interval in $[0, f_T)$ contains $x$.
By \cref{lem:tiles}\hyperref[lem:tiles:i]{(i)} and \Sref{1}, $S(x)$ is an independent set of $G$.
The function $S$ changes only at the at most $\sum_T |P_T| \le n$ interval endpoints.
Let $y^{(o)}$ give each non-empty set $I$ the total length of $\{x : S(x) = I\}$.
Then $y^{(o)}$ has at most $n$ non-zero entries and value at most $\max_T f_T \le \chif(G)$, and the sets containing a vertex $v \in P_T$ have total weight $\sum_{I \ni v} y^T_I \ge 1$.

By \cref{lem:tiles}\hyperref[lem:tiles:ii]{(ii)}, every vertex lies in an interior for at least $D - 2$ of the $D$ offsets.
So $y = \tfrac{1}{D-2}\sum_o y^{(o)}$ is a fractional coloring of $G$ of value at most $\tfrac{D}{D-2}\chif(G) \le (1+\delta)\chif(G)$, with at most $Dn$ non-zero entries.
\end{proof}

For unit disk and unit square graphs, \cref{lem:lp} also follows from~\cite[Theorem~5.2]{Jansen03} and~\cite[Theorem~2.10 and Section~4]{ErlebachJS05}.

\begin{lemma}[bounded differences~\cite{McDiarmid89}]\label{lem:mcd}
Let $Z_1, \dots, Z_N$ be independent random variables and $X = f(Z_1, \dots, Z_N)$, where changing any single argument changes $f$ by at most $1$.
Then $\Pr[X \ge \Exp X + t] \le \exp(-2t^2/N)$ for every $t \ge 0$.
\end{lemma}

\begin{lemma}[\cite{MoserT10}; symmetric form]\label{lem:mt}
Let $\mathcal Z$ be a finite set of mutually independent random variables, and let $\mathcal A$ be a finite set of events, each determined by a subset of $\mathcal Z$.
Suppose every event has probability at most $p$ and shares variables with at most $d \ge 1$ other events, and $e\,p\,(d+1) \le 1$.
Then the resampling algorithm finds an assignment of $\mathcal Z$ under which no event of $\mathcal A$ holds, after an expected number of at most $|\mathcal A|/d$ resampling steps.
\end{lemma}

Each resampling step needs only a way to sample the variables and to check whether a given event holds; both are polynomial in our setting.

\section{Asymptotic Approximation via LP Rounding}\label{sec:rounding}

This section proves \cref{lem:general}.
\Cref{sec:algorithm} describes the algorithm, which samples color classes from a fractional coloring independently in the tiles of $N$ shifted tilings and colors the remaining vertices with the sweep, and \cref{sec:analysis} analyzes it.

Heuristically, each round covers a vertex with probability at least $1/F$, where $F$ is the value of the fractional coloring, so after $sF$ rounds a vertex remains with probability about $e^{-s}$ and each clique $Q_j(v) \cap P$ keeps about $e^{-s}\omega \le e^{-s}F$ vertices.
The sweep then needs about $c\,e^{-s}F$ colors (\cref{lem:sweep}), for a total of $(s + c\,e^{-s})F$, minimized at $s = \ln c$.
Drawing independently per tile lets the Local Lemma bound all deviations at once by $O(\sqrt{F \log F})$.

\subsection{Algorithm}\label{sec:algorithm}

Fix a $c$-sweepable class, an input $G = (P,E)$ with $n = |P|$, and a fractional coloring $y$ of value $F$ with rational entries and polynomial support, as given by \cref{lem:lp}.
The algorithm has one integer parameter, the number of rounds $N \ge 3$, which is also the side of the tiles.

\begin{figure}[t]
\centering
\includegraphics{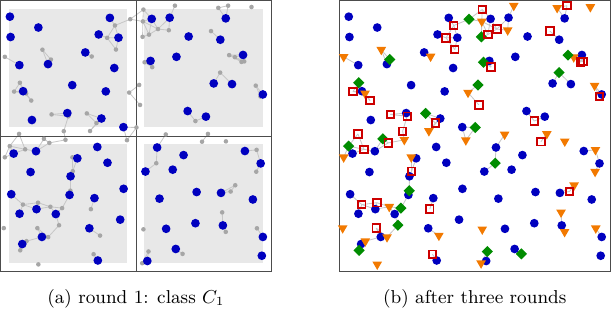}
\caption{Tiles of side $8$, a UDG on 150 random points (edges shaded), and random maximal independent sets in place of LP draws.
(a) Round~1: one independent set (large dots) per tile interior (shaded).
(b) After three rounds, each covered vertex has the color of its first class (blue circles, orange triangles, green diamonds); the residual (red squares) is colored by the sweep.}
\label{fig:example}
\end{figure}

\subparagraph*{Rounds.}
Round $k \in \{1, \dots, N\}$ uses the tiling $\mathcal T(N, k)$, so the tiling is shifted by one unit from each round to the next and every offset modulo $N$ is used exactly once.
\Cref{fig:example} shows the whole procedure on a small example.
For each round $k$ and each tile $T$ of $\mathcal T(N, k)$ whose interior contains a point of $P$ (at most $n$ tiles per round), independently draw a random independent set $Z_{k,T} \in \cI(G)$ with $\Pr[Z_{k,T} = I] = y_I / F$.

\subparagraph*{Color classes.}
For round $k$ let
\[
  C_k = \bigcup_T \bigl(Z_{k,T} \cap T^\circ\bigr).
\]
Give every vertex that lies in some $C_k$ the smallest such $k$ as its color.
Let $R$ be the set of remaining vertices (the \emph{residual}), and color $G[R]$ by the sweep with colors above $N$.

\subparagraph*{Local lemma.}
For $v \in P$ and $j \in [c]$ let $X_{v,j} = |Q_j(v) \cap P \cap R|$ and let $A_{v,j}$ be the event $X_{v,j} \ge \Exp X_{v,j} + t$, where the expectation is over independent draws and $t$ is fixed in \cref{sec:analysis}.
The draws $Z_{k,T}$ are resampled with the Moser--Tardos algorithm until no event $A_{v,j}$ holds.
The algorithm then outputs the coloring above.

\begin{lemma}\label{lem:proper}
Each $C_k$ is an independent set, so the output is a proper coloring.
\end{lemma}

\begin{proof}
Two vertices of $C_k$ in the same tile interior belong to the same independent set $Z_{k,T}$.
Two vertices in different interiors are at $L_\infty$ distance more than $1$ by \cref{lem:tiles}, so by \Sref{1} they are not adjacent.
Vertices colored in the first phase get the colors $1, \dots, N$, one class per round.
The residual gets colors above $N$ from a proper first-fit coloring.
\end{proof}

\subsection{Analysis}\label{sec:analysis}

We show that the algorithm runs in expected polynomial time and bound the number of colors it uses in terms of the LP value $F$.
For $v \in P$ let $\Gamma_v \subseteq [N]$ be the set of rounds $k$ in which $v$ lies in the interior of its tile.
The $N$ rounds use $N$ consecutive offsets, so $|\Gamma_v| \ge N - 2$ by \cref{lem:tiles}\hyperref[lem:tiles:ii]{(ii)}.

\begin{lemma}[expectation]\label{lem:expectation}
Let $N \ge sF + 2$ for some $s > 0$.
Then $\Pr[v \in R] \le e^{-s}$ for every $v$, and $\Exp X_{v,j} \le \omega\,e^{-s}$.
\end{lemma}

\begin{proof}
In a round $k \in \Gamma_v$, the vertex $v$ is covered if $v \in Z_{k,T}$ for its tile $T$, which happens with probability $\sum_{I \ni v} y_I / F \ge 1/F$.
These events are independent over $k$.
Hence
\[
  \Pr[v \in R] \le (1 - 1/F)^{|\Gamma_v|} \le \exp(-|\Gamma_v|/F) \le \exp(-(N-2)/F) \le e^{-s} .
\]
For the second claim, $P \cap Q_j(v)$ is a clique by \Sref{3} and so has at most $\omega$ vertices.
\end{proof}

\begin{lemma}[concentration]\label{lem:concentration}
$\Pr[A_{v,j}] \le \exp(-2t^2/N)$.
Moreover $A_{v,j}$ is determined by the draws $Z_{k,T}$ with $T^\circ \cap Q_j(v) \cap P \neq \emptyset$, of which there is at most one for each round $k$.
\end{lemma}

\begin{proof}
Let $Q = Q_j(v) \cap P$.
Since $\operatorname{diam}_\infty(Q_j(v)) \le 1$ and interiors of different tiles of one round are at $L_\infty$ distance more than $1$ (\cref{lem:tiles}\hyperref[lem:tiles:i]{(i)}), the set $Q$ meets at most one interior $T_k^\circ$ per round.
A vertex $u \in Q$ is covered if and only if $u \in Z_{k,T_k}$ for some round $k$ with $u \in T_k^\circ$.
So $X_{v,j} = |Q \cap R|$ is a function of the at most $N$ independent variables $Z_{k,T_k}$.
Changing one of them, say from $I$ to $I'$, changes the coverage in round $k$ only for the vertices of $I \cap Q \cap T_k^\circ$ and $I' \cap Q \cap T_k^\circ$.
Each of these sets has at most one vertex, because $Q$ is a clique and $I, I'$ are independent.
The first vertex may become uncovered and the second may become covered, so $X_{v,j}$ changes by at most $1$.
\Cref{lem:mcd} applies.
\end{proof}

\begin{lemma}[dependency]\label{lem:dependency}
Each event $A_{v,j}$ shares variables with fewer than $c(4N+5)^2\omega$ other events.
\end{lemma}

\begin{figure}[t]
\centering
\includegraphics{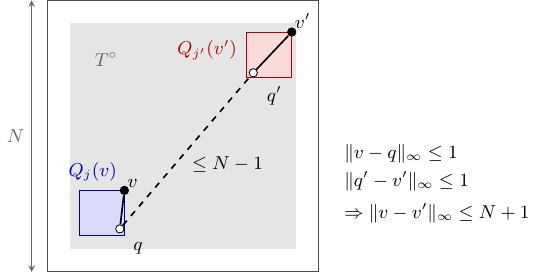}
\caption{Proof of \cref{lem:dependency}: if $Q_j(v)$ and $Q_{j'}(v')$ meet the same $T^\circ$, then $\inorm{v - v'} \le N + 1$.}
\label{fig:dependency}
\end{figure}

\begin{proof}
If $A_{v,j}$ and $A_{v',j'}$ share a variable $Z_{k,T}$, then there are $q \in Q_j(v) \cap T^\circ$ and $q' \in Q_{j'}(v') \cap T^\circ$.
Since $\operatorname{diam}_\infty(T^\circ) \le N - 1$ (\cref{fig:dependency}) and each $Q$-set contains its apex and has diameter at most $1$, we get $\inorm{v - v'} \le \inorm{v - q} + \inorm{q - q'} + \inorm{q' - v'} \le 1 + (N-1) + 1 = N + 1$.
The square $[x_v - N - 1, x_v + N + 1] \times [y_v - N - 1, y_v + N + 1]$ is covered by $(4N+5)^2$ half-open cells of side $1/2$.
Each cell has $L_\infty$-diameter at most $1/2$, so by \Sref{2} it contains at most $\omega$ points of $P$.
Each such $v'$ carries $c$ events, and this count includes $A_{v,j}$ itself.
\end{proof}

We now fix the parameters, choosing $t$ so that each event has probability at most $e^{-m}$ with $m \approx \ln d_0$, as \cref{lem:mt} needs.
Let $s > 0$ be a rational number and set
\[
  N = \lceil sF \rceil + 2, \qquad d_0 = c(4N+5)^2F, \qquad
  m = 1 + \bigl\lceil \tfrac{7}{10} \log_2 d_0 \bigr\rceil, \qquad t = \sqrt{\tfrac{mN}{2}} .
\]
All of these are computed exactly from the rational numbers $s$ and $F$: $m$ is the least integer with $2^{10(m-1)} \ge d_0^7$.
Since $\tfrac{7}{10} > \ln 2$, we have $\tfrac{7}{10}\log_2 d_0 \ge \ln d_0$, and therefore
\[
  1 + \ln d_0 \;\le\; m \;\le\; 2 + \tfrac{7}{10}\log_2 d_0 .
\]

\begin{lemma}\label{lem:round}
For every rational $s > 0$, the algorithm with these parameters terminates in expected polynomial time and uses at most
\[
  (s + c\,e^{-s})\,F + c\sqrt{mN/2} + 4 - c
\]
colors.
\end{lemma}

\begin{proof}
By \cref{lem:concentration,lem:dependency}, every event has probability at most $p = \exp(-2t^2/N) = e^{-m} \le 1/(e\,d_0)$ and, since the integer $c(4N+5)^2\omega$ is at most $d_0$, shares variables with at most $d := \lfloor d_0 \rfloor - 1$ others, where $d \ge 1$ since $d_0 \ge 2 \cdot 17^2$, so $e\,p\,(d+1) \le 1$.
There are $cn$ events and at most $nN$ variables, each sampled from an explicit distribution with polynomial support.
Each event can be evaluated exactly: $\Exp X_{v,j} = \sum_{u \in Q_j(v) \cap P} (1 - \sum_{I \ni u} y_I/F)^{|\Gamma_u|}$ is a rational number of polynomial size, and since $t^2 = mN/2$ is rational, the comparison with $t$ is exact after squaring.
By \cref{lem:mt}, the resampling stops after an expected polynomial number of steps with no event $A_{v,j}$ holding.

Then $X_{v,j} < \Exp X_{v,j} + t$ for all $v$ and $j$.
Every $v \in R$ belongs to $Q_j(v) \cap R$ for all $j$, so by \cref{lem:sweep} the residual uses at most $1 + \sum_j (X_{v,j} - 1)$ colors for the worst $v \in R$.
\Cref{lem:expectation} applies because $N \ge sF + 2$, so the total number of colors is at most
\[
  N + 1 - c + c\bigl(\omega\,e^{-s} + t\bigr) \;\le\; sF + 4 - c + c\,e^{-s}F + c\,t ,
\]
where we used $N \le sF + 3$ and $\omega \le F$.
\end{proof}

\begin{lemma}\label{lem:general}
Let $\mathcal G$ be a $c$-sweepable class.
For every $\varepsilon \in (0,1]$ there is a randomized algorithm with expected polynomial running time that colors every $G \in \mathcal G$ with at most $(1 + \ln c + \varepsilon)\chif(G) + K_\varepsilon$ colors, where $K_\varepsilon$ depends only on $c$ and $\varepsilon$.
\end{lemma}

\begin{proof}
Fix a rational number $s$ with $\ln c \le s \le \ln c + \varepsilon/4$, so that $s + c\,e^{-s} \le s + 1 \le 1 + \ln c + \varepsilon/4$.
Compute $y$ by \cref{lem:lp} with $F \le (1 + \delta)\chif$, where $\delta = \varepsilon/(2(2 + \ln c))$, and run the algorithm of \cref{sec:algorithm}.
In the bound of \cref{lem:round} we have $N \le sF + 3 \le (s+3)F$ and $d_0 \le c(4s+17)^2F^3$, so $m \le \gamma + 2.1\log_2 F \le \gamma + 3.1\ln F$ with $\gamma = 2 + \tfrac{7}{10}\log_2(c(4s+17)^2)$.
With $\lambda = c^2(s+3)/2$, the AM--GM inequality, and $\ln F \le F/\theta + \ln \theta - 1$ for $\theta = 49.6\lambda/\varepsilon^2$,
\begin{align*}
  c\sqrt{mN/2} &\;\le\; \sqrt{\lambda F(\gamma + 3.1\ln F)} \;\le\; \frac{\varepsilon F}{8} + \frac{2\lambda(\gamma + 3.1\ln F)}{\varepsilon}
  \;\le\; \frac{\varepsilon F}{4} + K', \\
  K' &\;=\; \frac{2\lambda}{\varepsilon}\Bigl(\gamma + 3.1\ln\frac{50\lambda}{\varepsilon^2}\Bigr),
\end{align*}
an explicit constant that depends only on $c$ and $\varepsilon$.
As $4 - c \le 2$, the algorithm uses at most
\[
  (1 + \ln c + \varepsilon/2)F + K' + 2 \;\le\; (1 + \ln c + \varepsilon/2)(1+\delta)\chif + K' + 2
  \;\le\; (1 + \ln c + \varepsilon)\chif + K' + 2
\]
colors, because $(1 + \ln c + \varepsilon/2)\delta \le (2 + \ln c)\delta = \varepsilon/2$.
\end{proof}

The bound of \cref{lem:round} holds for every $F$, so no separate treatment of small $F$ is needed.
As $s$ tends to $\ln c$ it becomes $(1 + \ln c)F + O(\sqrt{F \log F})$; the $\varepsilon$ of \cref{lem:general} pays only for the LP, for a rational $s$, and for this lower-order term.
For unit disk graphs, \cref{lem:general} gives the constant $1 + \ln 3$, which \cref{sec:reuse} lowers to that of \cref{thm:frac}.

\section{A 2.1-Approximation Algorithm}\label{sec:absolute}

This section removes the additive constant of \cref{lem:general} for classes with $c \ge 3$ (\cref{lem:absolute}), which gives a $2.1$-approximation for unit disk graphs and for translates of a convex polygon.
The additive constant $K_\varepsilon$ of \cref{lem:general} is at most $\varepsilon\chif$ once $\chif \ge K_\varepsilon/\varepsilon$.
So it matters only when $\chif$, and hence $\omega \le \chif$, is below a constant, and it remains to color graphs of bounded clique number.
Inside a vertical slab of bounded width, sorted by $y$, such a graph has bounded bandwidth, so it can be colored optimally in polynomial time~\cite{ArnborgP89}.

\begin{lemma}[coloring a slab]\label{lem:slabdp}
Let $\mathcal G$ be a $c$-sweepable class and let $W \ge 1$ be an integer.
There is an algorithm that, given an integer $C \ge 1$ and a graph $G \in \mathcal G$ on $n$ points with $\omega(G) \le C$ that lie in a vertical slab $a \le x < a + W$, computes an optimal coloring of $G$ in $O(n \log n) + 2^{O(WC \log (WC))} n$ time.
\end{lemma}

\begin{figure}[t]
\centering
\includegraphics{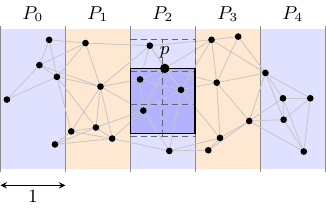}
\caption{\Cref{lem:slabdp,lem:slabs}: slabs alternate palettes; earlier neighbors of $p$ lie in six cells.}
\label{fig:slabs}
\end{figure}

\begin{proof}
Order the points by $y$-coordinate, breaking ties arbitrarily.
If $q$ precedes $p$ and $qp$ is an edge, then by \Sref{1} the points $q$, $p$, and all points between them lie in $[a, a+W) \times [y_p - 1, y_p]$, which is covered by $6W$ half-open cells of side $1/2$ (\cref{fig:slabs}) and so contains at most $b = 6WC$ points by \Sref{2}; so every earlier neighbor of $p$ is among the $b - 1$ points preceding it.
Hence first-fit in this order uses at most $b$ colors, and for each $k \le b$, dynamic programming along the order that keeps the $k$-colorings of the last $b - 1$ points that extend to all earlier points decides whether $G$ is $k$-colorable; the least such $k$ gives an optimal coloring.
Sorting takes $O(n \log n)$ time, and for each $k$ the dynamic program keeps at most $k^{b-1}$ colorings and tries $k$ colors at each point, each checked against the $b - 1$ predecessors with its adjacency list, of length less than $25C$ by \Sref{1} and \Sref{2}, so all $k \le b$ take $2^{O(WC \log (WC))} n$ time.
\end{proof}

Two alternating palettes on slabs of width $1$ then lose only a factor $2$.

\begin{lemma}[slabs]\label{lem:slabs}
Let $\mathcal G$ be a $c$-sweepable class.
There is an algorithm that, given an integer $C \ge 1$ and a graph $G \in \mathcal G$ on $n$ points with $\omega(G) \le C$, colors $G$ with at most $2\chi(G)$ colors in $O(n \log n) + 2^{O(C \log C)} n$ time.
\end{lemma}

\begin{proof}
By \Sref{1}, no edge joins the slabs $P_i = \{p \in P : i \le x_p < i + 1\}$ and $P_{i'}$ if $|i - i'| \ge 2$.
So coloring each $G[P_i]$ optimally by \cref{lem:slabdp} with $W = 1$, from one palette if $i$ is even and another if $i$ is odd, gives at most $2\max_i \chi(G[P_i]) \le 2\chi(G)$ colors.
\end{proof}

\begin{lemma}\label{lem:absolute}
Let $\mathcal G$ be a $c$-sweepable class with $c \ge 3$.
For every $\varepsilon \in (0,1]$ there is a randomized algorithm with expected polynomial running time that colors every $G \in \mathcal G$ with at most $(1 + \ln c + \varepsilon)\chi(G)$ colors.
\end{lemma}

\begin{proof}
Let $K = K_{\varepsilon/2} = K' + 2$ be the explicit constant of \cref{lem:general} and let $C \ge 4K/\varepsilon$ be an integer.
Compute a fractional coloring of value $F \le 2\chif$ by \cref{lem:lp}.
If $F \ge C$, then $\chif \ge 2K/\varepsilon$, and the algorithm of \cref{lem:general} uses at most $(1 + \ln c + \varepsilon/2)\chif + K \le (1 + \ln c + \varepsilon)\chif \le (1 + \ln c + \varepsilon)\chi$ colors.
Otherwise $\omega \le \chif \le F < C$, and \cref{lem:slabs} gives at most $2\chi \le (1 + \ln c)\chi$ colors, because $\ln c \ge \ln 3 > 1$.
\end{proof}

By \cref{lem:classes}, \cref{lem:absolute} with $c = 3$ and $\varepsilon = 1/1000$ colors every unit disk graph with at most $(1 + \ln 3 + 1/1000)\,\chi < 2.0997\,\chi$ colors.

\subsection{Translates of a Convex Polygon}\label{sec:translates}

This subsection shows that intersection graphs of translates of a convex polygon are $3$-sweepable after a linear change of coordinates, which extends \cref{lem:general,lem:absolute,thm:reuse} with $c = 3$ to them and proves \cref{thm:translates}.
Let $M \subset \R^2$ be a convex polygon with non-empty interior, and let $P \subset \R^2$ be finite.
The intersection graph of the translates $M + p$, $p \in P$, has an edge $pq$ if and only if $p - q$ lies in $B := M - M = \{x - y : x, y \in M\}$.
The set $B$ is a convex polygon that is symmetric about the origin, and unit disk and unit square graphs are defined in the same way with a disk or a square as $B$.
A sweep gives $\chi \le 3\omega - 2$ for these graphs~\cite{KimKN04}, which is a $3$-approximation; we are not aware of a smaller ratio.
If $M$ is a parallelogram, a linear map turns the graphs into unit square graphs.

\begin{lemma}[{implicit in~\cite[Theorem~8]{KimKN04}}]\label{lem:translates}
After a linear change of coordinates that depends only on $M$, the intersection graphs of translates of $M$ form a $3$-sweepable class.
\end{lemma}

\begin{figure}[t]
\centering
\includegraphics{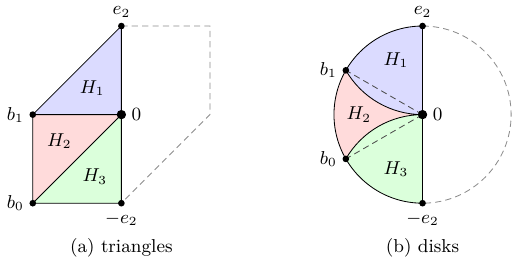}
\caption{The sets $H_j$, with $Q_j(v) = v + H_j$, and the points $b_0$, $b_1$ of the proof of \cref{lem:translates}.
(a) For the triangle $M$ with vertices $(0,0)$, $(1,0)$, $(1,1)$, the hexagon $B$ is that of \cref{fig:shapes}(a), and the cones $\angle(e_2, b_1)$, $\angle(b_1, b_0)$, $\angle(b_0, -e_2)$ are the three triangles.
(b) For disks the sets $H_j$ are curved, unlike the sectors of \cref{fig:tiles}(c), and $H_2$ lies strictly inside the cone $\angle(b_1, b_0)$ (dashed rays).}
\label{fig:translates}
\end{figure}

\begin{proof}
Let $e_1, e_2$ be vertices of $B$ that maximize $|\det(e_1, e_2)|$; the maximum of $|\det(x, y)|$ over $x, y \in B$ is attained at vertices.
Apply the linear map that takes $e_1, e_2$ to $(1,0), (0,1)$.
In the new coordinates, every $w \in B$ has $|x_w| = |\det(w, e_2)| / |\det(e_1, e_2)| \le 1$ and likewise $|y_w| \le 1$, so $B \subseteq [-1,1]^2$, which is \Sref{1}.
Moreover $B$ contains the convex hull of $\pm e_1, \pm e_2$, hence the square $[-\tfrac12, \tfrac12]^2$, which is \Sref{2}.

For \Sref{3}, let $\prec$ be the lexicographic order on $(x,y)$, as in \cref{lem:classes}.
Let $H = \{w \in B : x_w \le 0\}$, so that every neighbor $u \prec v$ of $v$ lies in $v + H$, and define
\begin{align*}
  Q_1(v) &= v + \{w \in H : w - e_2 \in B\}, \qquad Q_3(v) = v + \{w \in H : w + e_2 \in B\}, \\
  Q_2(v) &= v + \bigl(\{0\} \cup \{w \in H : w + e_2 \notin B,\ w - e_2 \notin B\}\bigr).
\end{align*}
These sets contain $v$ and cover $v + H$ (\cref{fig:translates}), and membership is decided by linear inequalities.
Let $H_1, H_2, H_3 \subseteq H$ be the sets with $Q_j(v) = v + H_j$.
Since $B \subseteq [-1,1]^2$, it suffices to show that $w - w' \in B$ whenever $w, w' \in H_j$ for some $j$: then $P \cap Q_j(v)$ is a clique and $\operatorname{diam}_\infty(Q_j(v)) \le 1$.
This is the content of Claims~3.1--3.3 in the proof of~\cite[Theorem~8]{KimKN04}; we give a self-contained proof.

Let $\|x\| = \min\{\lambda \ge 0 : x \in \lambda B\}$ be the norm with unit ball $B$, so that $w - w' \in B$ if and only if $\|w - w'\| \le 1$, and $x \mapsto \|x - q\|$ is convex for every $q$.
The vertices $\pm e_1$ and $\pm e_2$ of $B$ have norm $1$, and $0$ is an interior point of $B$.
We use one inequality: if two segments $[p, p']$ and $[q, q']$ meet at a point $z$, then the triangle inequality at $z$ gives
\begin{align*}
  \|p - p'\| + \|q - q'\| &\;=\; \|p - z\| + \|z - p'\| + \|q - z\| + \|z - q'\| \\
  &\;\ge\; \max\bigl\{\|p - q\| + \|p' - q'\|,\ \|p - q'\| + \|q - p'\|\bigr\}. \tag{$\ast$}
\end{align*}

\emph{The set $H_1$.}
Let $w, w' \in H_1$.
The claim is immediate if $w$ or $w'$ is $0$ or $e_2$, so assume not.
Since $w - e_2 \in B \subseteq [-1,1]^2$, we have $y_w \ge 0$, and likewise $y_{w'} \ge 0$, so the four points $0, e_2, w, w'$ lie in the strip $\{x \le 0,\ 0 \le y \le 1\}$, of which $0$ and $e_2$ are corners.
A corner of the strip lies in the convex hull of other points of the strip only if it is one of them, so neither $0$ nor $e_2$ lies in the convex hull of the other three points.
If $w$ lies in the convex hull of $0$, $e_2$, $w'$, then by convexity $\|w - w'\| \le \max\{\|0 - w'\|, \|e_2 - w'\|, 0\} \le 1$, and likewise if $w'$ lies in the convex hull of $0$, $e_2$, $w$.
Otherwise the four points are the vertices of a convex quadrilateral.
Its diagonals cross, so $0$ and $e_2$ are not opposite vertices, as $w$ and $w'$ would then lie strictly on opposite sides of the $y$-axis.
So the diagonals are $[0, r]$ and $[e_2, r']$ with $\{r, r'\} = \{w, w'\}$, and $(\ast)$ gives $2 \ge \|r\| + \|e_2 - r'\| \ge \|e_2\| + \|r - r'\| = 1 + \|w - w'\|$.
The same argument with $-e_2$ in place of $e_2$ and the strip $\{x \le 0,\ -1 \le y \le 0\}$ applies to $H_3$.

\emph{The set $H_2$.}
The set $B \cap (B - e_2)$ contains $(-\tfrac12, -\tfrac12)$; let $b_0$ be one of its points with the least $x$-coordinate, so $x_{b_0} \le -\tfrac12$, and let $b_1 = b_0 + e_2$.
We first show that $b_0$ and $b_1$ lie on the boundary of $B$.
For $x \in [-1, 1]$ let $\ell(x)$ be the length of the vertical chord $\{y : (x, y) \in B\}$ of $B$; it is concave by convexity of $B$, and $\ell(x) < 1$ for $x < x_{b_0}$ by the choice of $b_0$.
The chord at $x_{b_0}$ contains $b_0$ and $b_1$, which are at vertical distance $1$.
If $x_{b_0} = -1$, then $b_0$ and $b_1$ are on the boundary of $B \subseteq [-1,1]^2$.
Otherwise $\ell$ is continuous at the interior point $x_{b_0}$, so $\ell(x_{b_0}) \le 1$, the chord at $x_{b_0}$ is the segment $[b_0, b_1]$, and its endpoints lie on the boundary of $B$.
Hence $\|b_0\| = \|b_1\| = \|b_1 - b_0\| = 1$.

For $p, q \in \R^2$ let $\angle(p, q) = \{\lambda p + \mu q : \lambda, \mu \ge 0\}$.
As $x_{b_1} = x_{b_0} < 0$ and $y_{b_1} > y_{b_0}$, the half-plane $\{x \le 0\}$ is the union of $\angle(e_2, b_1)$, $\angle(b_1, b_0)$, and $\angle(b_0, -e_2)$.
We show that $B \cap \angle(e_2, b_1) \subseteq H_1$.
Let $w \in B \cap \angle(e_2, b_1)$.
If $w$ lies in the triangle with vertices $0$, $e_2$, $b_1$, then $\|w - e_2\| \le \max\{\|e_2\|, 0, \|b_1 - e_2\|\} = 1$ by convexity.
Otherwise the segment $[0, w]$ crosses $[e_2, b_1]$, and $(\ast)$ gives $\|w\| + \|e_2 - b_1\| \ge \|b_1\| + \|e_2 - w\|$, so $\|w - e_2\| \le 1$.
In the same way, using $\|b_0\| = \|{-e_2} - b_0\| = 1$, we get $B \cap \angle(b_0, -e_2) \subseteq H_3$.
Therefore $H_2 \subseteq B \cap \angle(b_1, b_0)$.

It remains to show that $\|u - u'\| \le 1$ for $u, u' \in B \cap \angle(b_1, b_0)$.
This set is convex and $\|u - u'\|$ is convex in $u$ and in $u'$, so we may assume that $u$ and $u'$ are extreme points of the set.
Since $\|u\| \le 1$ for every $u \in B$, we may also assume that $u, u' \ne 0$, so $u$ and $u'$ lie on the boundary of $B$, and after exchanging them, $u \in \angle(b_1, b_0)$ and $u' \in \angle(u, b_0)$.
Let $p \in B \cap \angle(b_1, b_0)$ and let $q \in \angle(p, b_0)$ lie on the boundary of $B$.
The ray from $0$ through $q$ meets the segment $[p, b_0] \subseteq B$, and since $0$ is an interior point of $B$, the points of $B$ on this ray form the segment $[0, q]$.
So $[0, q]$ meets $[p, b_0]$, and $(\ast)$ gives $\|q\| + \|p - b_0\| \ge \max\{\|p\| + \|q - b_0\|,\ \|b_0\| + \|p - q\|\}$.
With $(p, q) = (b_1, u)$ this gives $\|u - b_0\| \le \|b_1 - b_0\| = 1$, and with $(p, q) = (u, u')$ it gives $\|u - u'\| \le \|u - b_0\| \le 1$.
\end{proof}

\begin{figure}[t]
\centering
\includegraphics{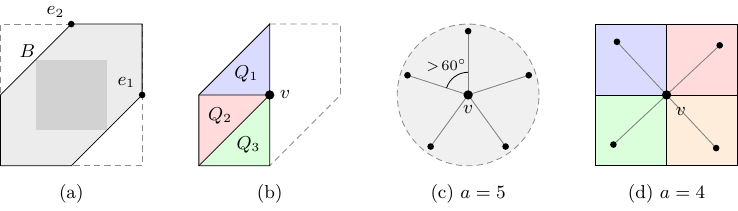}
\caption{(a) $B = M - M$ for a triangle $M$, between $[-\tfrac12,\tfrac12]^2$ and $[-1,1]^2$.
(b) The cliques $Q_j(v)$.
Independent neighbors of $v$: (c) disks, ${>}\,60^\circ$ apart; (d) squares, one per quadrant.}
\label{fig:shapes}
\end{figure}

By \cref{lem:translates}, \cref{lem:absolute} with $c = 3$ and $\varepsilon = 1/1000$ colors every intersection graph of translates of a convex polygon with at most $(1 + \ln 3 + 1/1000)\,\chi < 2.0997\,\chi$ colors.

\section{A 2.05-Approximation Algorithm}\label{sec:reuse}

This section lowers the leading constant $1 + \ln c$ of \cref{lem:general}.
The algorithm of \cref{sec:algorithm} colors the residual with colors above $N$.
We now color $G[R]$ instead by first-fit in the order $\prec$ of \Sref{3} in \cref{def:sweepable} (for unit disks, the lexicographic order on $(x,y)$), giving each $v \in R$ the smallest positive integer that is not the color of an already colored neighbor, including the neighbors colored in the first phase.
So a residual vertex may receive a color $k \le N$ that none of its neighbors received.
The algorithm is otherwise unchanged, except that the Local Lemma step also controls the number of such colors.

\subparagraph*{Neighborhood independence.}
A class of graphs has \emph{neighborhood independence} at most $a$ if, for every graph $G$ in the class and every vertex $v$, every independent set of $G$ contained in the neighborhood $N_G(v)$ has at most $a$ vertices.
Unit disk graphs have neighborhood independence at most $5$ (\cref{fig:shapes}(c)).
Indeed, a neighbor at the same point as $v$ is adjacent to every other neighbor of $v$, and two neighbors $u, w$ of $v$ at other points that make an angle of at most $60^\circ$ at $v$ satisfy $\|u - w\|_2 \le \max\{\|u - v\|_2, \|w - v\|_2\} \le 1$.
So the vertices of an independent set in $N_G(v)$ with at least two vertices make pairwise angles larger than $60^\circ$ at $v$, and there are at most $5$ of them.
Intersection graphs of translates of a convex polygon have neighborhood independence at most $5$ as well, as we show at the end of this section.

\subparagraph*{Free colors.}
Fix a $c$-sweepable class with neighborhood independence at most $a$.
For $v \in P$ let $U(v)$ be the number of colors $k \in [N]$ that no neighbor of $v$ receives in the first phase; we call them \emph{free} for $v$.

\begin{lemma}\label{lem:reuseff}
The coloring is proper and uses at most $N + \max\bigl\{0,\ \max_{v \in R}\bigl(1 + \sum_{j=1}^c (X_{v,j} - 1) - U(v)\bigr)\bigr\}$ colors.
\end{lemma}

\begin{proof}
The first phase is proper by \cref{lem:proper}, and first-fit keeps the coloring proper.
When $v \in R$ is colored, its neighbors colored in the first phase carry the $N - U(v)$ colors of $[N]$ that are not free for $v$.
Its neighbors in $R$ colored before it lie in $\bigcup_j (Q_j(v) \cap R) \setminus \{v\}$ by \Sref{3}, so there are at most $\sum_j (X_{v,j} - 1)$ of them.
Hence at most $N - U(v) + \sum_j (X_{v,j} - 1)$ colors are excluded for $v$, and $v$ receives a color of at most one more.
\end{proof}

Color $k \ge 4$ is free for $v$ if and only if the at most $a$ neighbors of $v$ in $C_k$, which depend only on round $k$, are all covered in earlier rounds.
Each of them is covered before round $k$ with probability at least $1 - e^{-(k-3)/F}$, and a Poisson coupling, which costs the last term below, and Harris' inequality give the product bound.

\begin{lemma}\label{lem:freeexp}
For every $v \in P$,
\[
  \Exp U(v) \;\ge\; \sum_{k=4}^{N} \bigl(1 - e^{-(k-3)/F}\bigr)^a - \frac{a^2N^2}{2F^2} .
\]
\end{lemma}

\begin{proof}
Fix a round $k \ge 4$ and let $S = N_G(v) \cap C_k$.
A neighbor of $v$ receives color $k$ if and only if it lies in $C_k$ and in no earlier $C_j$, so $k$ is free for $v$ if and only if every vertex of $S$ lies in $C_1 \cup \dots \cup C_{k-1}$.
The set $S$ is independent, so it has at most $a$ vertices, and it depends only on the draws of round $k$, which are independent of the draws of the earlier rounds.
It therefore suffices to show, for every fixed independent set $S \subseteq N_G(v)$ with $h \le a$ vertices, that $S \subseteq C_1 \cup \dots \cup C_{k-1}$ with probability at least $(1 - e^{-(k-3)/F})^h - (k-1)h^2/F^2$.
This is at least $(1 - e^{-(k-3)/F})^a - (k-1)a^2/F^2$, and summing over $k$ with $\sum_k (k - 1) \le N^2/2$ gives the lemma.

As in the proof of \cref{lem:expectation}, a vertex $u$ lies in $C_j$ with probability $p_{u,j} \ge 1/F$ for every $j \in \Gamma_u$.
For $j < k$, let $D_j$ contain each $u \in S \cap C_j$ with $j \in \Gamma_u$ independently with probability $1/(F p_{u,j})$.
Then $D_1, \dots, D_{k-1}$ are independent, $\Pr[u \in D_j] = 1/F$ if $j \in \Gamma_u$ and $0$ otherwise, $q_j := \Pr[D_j \neq \emptyset] \le h/F$, and it suffices to bound the probability that $S \subseteq D_1 \cup \dots \cup D_{k-1}$.
For $\emptyset \neq J \subseteq S$ let $r_{j,J} = \Pr[D_j = J]$, and let $Y_{j,J}$ be independent Poisson variables with means $r_{j,J}$.
For each $j$, the sum $\sum_J Y_{j,J}$ is Poisson with mean $q_j$, and given that it is $1$, the unique $J$ with $Y_{j,J} = 1$ has the law of $D_j$ given $D_j \neq \emptyset$.
The total variation distance between the Bernoulli and the Poisson distribution with mean $q$ is $q(1 - e^{-q}) \le q^2$.
So $D_j$ and $(Y_{j,J})_J$ can be coupled, independently over $j$, such that with probability at least $1 - h^2/F^2$ either $D_j = \emptyset$ and all $Y_{j,J}$ vanish, or $D_j = J$, $Y_{j,J} = 1$, and all other $Y_{j,J'}$ vanish.

Outside an event of probability at most $(k-1)h^2/F^2$, therefore, $S \subseteq D_1 \cup \dots \cup D_{k-1}$ holds if and only if, for every $u \in S$, some $Y_{j,J}$ with $j < k$ and $J \ni u$ is positive.
These $h$ events are increasing in the independent indicator variables $[Y_{j,J} \ge 1]$, so by Harris' inequality~\cite{Harris60} their intersection has probability at least the product of their probabilities.
The event for $u$ fails with probability $\prod_{j < k}\prod_{J \ni u} e^{-r_{j,J}} = \exp\bigl(-\sum_{j<k} \Pr[u \in D_j]\bigr) = \exp(-|\Gamma_u \cap [k-1]|/F)$, which is at most $e^{-(k-3)/F}$ because $|\Gamma_u \cap [k-1]| \ge k - 3$ by \cref{lem:tiles}\hyperref[lem:tiles:ii]{(ii)}.
\end{proof}

\begin{lemma}\label{lem:freeconc}
$U(v)$ is determined by the draws $Z_{k,T}$ with $T^\circ \cap N_G(v) \neq \emptyset$, of which there are at most four for each round $k$, and $\Pr[U(v) \le \Exp U(v) - t'] \le \exp\bigl(-t'^2/(2(2a+1)^2N)\bigr)$ for every $t' \ge 0$.
\end{lemma}

\begin{proof}
By \Sref{1}, $N_G(v)$ lies in $v + [-1,1]^2$.
In each coordinate, the interiors of a tiling $\mathcal T(N,k)$ are intervals of length $N - 1 \ge 2$ separated by gaps of length $1$, so an interval of length $2$ meets at most two of them, and $v + [-1,1]^2$ meets at most four interiors.
The first-phase color of a neighbor $u$ of $v$ depends only on the draws for the tiles whose interiors contain $u$.
Changing one draw $Z_{k,T}$ from $I$ to $I'$ changes the first-phase color only of vertices in $(I \cup I') \cap T^\circ$, and at most $2a$ of them are neighbors of $v$, because $I \cap N_G(v)$ and $I' \cap N_G(v)$ are independent.
Each of these changes its color from or to $k$, so the set of neighbors of $v$ with a given color changes for at most $2a + 1$ colors, and $U(v)$ changes by at most $2a + 1$.
\Cref{lem:mcd} applied to $-U(v)/(2a+1)$, a function of at most $4N$ independent variables, gives the bound.
\end{proof}

\subparagraph*{Parameters.}
With $s$, $N$, $m$, and $t$ as in \cref{sec:analysis}, but $d_0 = (c+1)(4N+5)^2F$, let
\[
  \beta = \sum_{k=4}^{N} \bigl(1 - (1 + 1/F)^{3-k}\bigr)^a - \frac{a^2N^2}{2F^2}, \qquad t' = (2a+1)\sqrt{2mN} .
\]
Since $(1 + 1/F)^{k-3} \le e^{(k-3)/F}$, we have $\beta \le \Exp U(v)$ by \cref{lem:freeexp}.
For every $v \in P$ let $E_v$ be the event $U(v) \le \beta - t'$; by \cref{lem:freeconc} it has probability at most $e^{-m}$.
The Moser--Tardos algorithm now resamples the draws until none of the events $A_{v,j}$ and $E_v$ holds.
For $\Phi_a(s) = \int_0^s (1 - e^{-x})^a\,dx$ let
\[
  \kappa_{c,a}(s) = s + \max\bigl\{0,\ c\,e^{-s} - \Phi_a(s)\bigr\} .
\]

\begin{lemma}\label{lem:reuseround}
For every rational $s \in (0, 2]$ and $F \ge 3$, the algorithm with these parameters terminates in expected polynomial time and uses at most
\[
  \kappa_{c,a}(s)\,F + c\sqrt{mN/2} + (2a+1)\sqrt{2mN} + 4 + \tfrac12(a + a^2)(s + 3/F)^2
\]
colors.
\end{lemma}

\begin{proof}
The event $E_v$ is determined by the draws for tiles whose interiors meet $N_G(v)$, and every point of $N_G(v)$ is within $L_\infty$ distance $1$ of $v$, so the proof of \cref{lem:dependency} shows that each event shares variables with fewer than $(c+1)(4N+5)^2\omega \le d_0$ others.
As in the proof of \cref{lem:round}, $e\,p\,(d+1) \le 1$ with $p = e^{-m}$, and each event $E_v$ is evaluated exactly, since $\beta$ and $t'^2$ are rational.
So the resampling stops after an expected polynomial number of steps, with $X_{v,j} < \Exp X_{v,j} + t \le \omega e^{-s} + t$ and $U(v) > \beta - t'$ for all $v$ and $j$.
By \cref{lem:reuseff} and $N \le sF + 3$, $\omega \le F$, the number of colors is at most
\[
  sF + 3 + \max\bigl\{0,\ c\,e^{-s}F - \beta\bigr\} + c\,t + t' .
\]
It remains to bound $\beta$ from below.
From $\ln(1 + u) \ge u - u^2/2$ and $e^y \le 1 + 2y$ for $y \in [0,1]$ we get $(1 + 1/F)^{-x} \le e^{-x/F} + x/F^2$ for $0 \le x \le 2F^2$, which covers $x \le N \le sF + 3$.
As $z \mapsto z^a$ is $a$-Lipschitz on $[0,1]$, each term of $\beta$ is at least $(1 - e^{-(k-3)/F})^a - a(k-3)/F^2$, and these corrections sum to at most $aN^2/(2F^2)$.
The function $x \mapsto (1 - e^{-x})^a$ is nondecreasing and at most $1$, and $N - 3 \ge sF - 1$, so
\[
  \beta \;\ge\; F\,\Phi_a\Bigl(\frac{N-3}{F}\Bigr) - \frac{(a + a^2)N^2}{2F^2} \;\ge\; F\,\Phi_a(s) - 1 - \tfrac12(a + a^2)(s + 3/F)^2 .
\]
Hence $\max\{0, c\,e^{-s}F - \beta\} \le \max\{0, c\,e^{-s} - \Phi_a(s)\}F + 1 + \tfrac12(a + a^2)(s + 3/F)^2$, which gives the bound.
\end{proof}

\begin{theorem}\label{thm:reuse}
Let $\mathcal G$ be a $c$-sweepable class with neighborhood independence at most $a$, and let $s \in (0,2]$ be rational.
For every $\varepsilon \in (0,1]$ there is a randomized algorithm with expected polynomial running time that colors every $G \in \mathcal G$ with at most $(\kappa_{c,a}(s) + \varepsilon)\chif(G) + K$ colors, where $K$ depends only on $c$, $a$, and $\varepsilon$.
\end{theorem}

\begin{proof}
Compute $y$ by \cref{lem:lp} with $F \le (1 + \delta)\chif$, where $\delta = \varepsilon/(2(c+3))$.
If $F < 3$, then $\omega \le 2$ and the sweep uses at most $c + 1$ colors.
Otherwise run the algorithm of \cref{lem:reuseround}.
As in the proof of \cref{lem:general}, since $s \le 2$, $N \le sF + 3 \le 5F$ and $d_0 = (c+1)(4N+5)^2F \le 625(c+1)F^3$, so $m \le \gamma + 3.1\ln F$ with $\gamma = 2 + \tfrac{7}{10}\log_2(625(c+1))$.
The two square-root terms of \cref{lem:reuseround} sum to $\bigl(c/\sqrt2 + \sqrt2(2a+1)\bigr)\sqrt{mN} \le \sqrt{\lambda F m}$ with $\lambda = 5(c + 4a + 2)^2/2$.
The AM--GM step and the tangent-line bound of the proof of \cref{lem:general}, with this $\lambda$ and $\gamma$, give $\sqrt{\lambda F m} \le \varepsilon F/4 + K'$ with $K' = \tfrac{2\lambda}{\varepsilon}\bigl(\gamma + 3.1\ln\tfrac{50\lambda}{\varepsilon^2}\bigr)$.
Since $s \le 2$ and $F \ge 3$, the remaining terms are at most $4 + \tfrac92(a + a^2)$.
So the algorithm uses at most $(\kappa_{c,a}(s) + \varepsilon/4)F + K'' \le (\kappa_{c,a}(s) + \varepsilon)\chif + K''$ colors, where $K'' = K' + 4 + \tfrac92(a+a^2)$, because $\kappa_{c,a}(s) + 1 \le s + c + 1 \le c + 3$ gives $(\kappa_{c,a}(s) + \varepsilon/4)\delta \le \varepsilon/2$.
The theorem holds with $K = \max\{K'', c + 1\}$.
\end{proof}

If $c\,e^{-s} \ge \Phi_a(s)$, substituting $y = e^{-x}$ gives, with $x_0 = e^{-s}$ and the harmonic number $H_a$,
\[
  \kappa_{c,a}(s) = c\,x_0 + \int_{x_0}^1 \frac{1 - (1 - y)^a}{y}\,dy
  = c\,x_0 + H_a - \sum_{i=1}^a \frac{(-1)^{i+1}}{i}\binom{a}{i}x_0^i .
\]
Without reuse the integrand is $1/y$ and the bound is $c\,x_0 + s$, as in \cref{lem:general}; reuse saves $\int_{x_0}^1 (1-y)^a/y\,dy$.
For unit disk graphs ($c = 3$, $a = 5$) and $s = 27/20$, we get $c\,e^{-s} \ge \Phi_5(s)$ and $\kappa_{3,5}(s) < 2.04822$, while $1 + \ln 3 > 2.0986$.
This value is within $10^{-4}$ of the minimum over $s$.
With $\varepsilon = 10^{-6}$, \cref{thm:reuse} therefore gives \cref{thm:frac}.

\begin{proof}[Proof of \cref{thm:main}]
Follow the proof of \cref{lem:absolute} with \cref{thm:reuse} for $c = 3$, $a = 5$, $s = 27/20$, and $\varepsilon = 10^{-6}$ in place of \cref{lem:general}.
If $F \ge C$, this uses at most $(\kappa_{3,5}(27/20) + 10^{-6})\chi < 2.0483\,\chi$ colors.
Otherwise \cref{lem:slabs} uses at most $2\chi$ colors, and $2 < \kappa_{3,5}(27/20)$.
\end{proof}

\subparagraph*{Translates of a convex polygon.}
The proofs of \cref{thm:frac,thm:main} use only that the class is $3$-sweepable with neighborhood independence at most $5$.
Intersection graphs of translates of a convex polygon $M$ are $3$-sweepable by \cref{lem:translates}, and the next lemma gives the second property.
Intuitively, in the norm with unit ball $M - M$, moving the points of an independent set in $N_G(v)$ away from $v$ to distance exactly $1$ keeps them more than $1$ apart, so their translates of $M$ touch $M + v$ and are pairwise disjoint, and there are at most $5$ such translates~\cite{DoyleLR92}.

\begin{lemma}\label{lem:translatesindep}
Intersection graphs of translates of a convex polygon have neighborhood independence at most $5$.
\end{lemma}

\begin{proof}
Let $\|x\| = \min\{\lambda \ge 0 : x \in \lambda B\}$ be the norm with unit ball $B = M - M$, so that $pq$ is an edge if and only if $\|p - q\| \le 1$, and translates $M + p$ and $M + q$ are disjoint if and only if $\|p - q\| > 1$ and touch, that is, intersect with disjoint interiors, if and only if $\|p - q\| = 1$.
Let $S$ be an independent set in the neighborhood of a vertex $v$ with $|S| \ge 2$.
No $u \in S$ lies at the point $v$, since such a $u$ would be adjacent to every other neighbor of $v$.
So every $u \in S$ can be written as $u = v + \sigma_u \xi_u$ with $0 < \sigma_u \le 1$ and $\|\xi_u\| = 1$.
For $u, w \in S$ the function $g(\sigma, \tau) = \|\sigma \xi_u - \tau \xi_w\|$ is convex on $[0,1]^2$, so it attains its maximum at a corner, and
\[
  1 < \|u - w\| = g(\sigma_u, \sigma_w) \le \max\{g(0,0),\, g(1,0),\, g(0,1),\, g(1,1)\} = \max\{0,\, 1,\, 1,\, \|\xi_u - \xi_w\|\} .
\]
Hence $\|\xi_u - \xi_w\| > 1$, and the translates $M + v + \xi_u$, $u \in S$, are pairwise disjoint and all touch $M + v$.
They are therefore distinct, and after translating by $-v$ they form a \emph{strict Hadwiger family} of $M$, that is, pairwise disjoint translates of $M$ that all touch $M$.
The same vectors give a strict Hadwiger family of the centrally symmetric body $\tfrac12 B$, since $p + \tfrac12 B$ and $q + \tfrac12 B$ are disjoint or touch exactly when $M + p$ and $M + q$ are.
So if $M$ is not a parallelogram, the family has at most $5$ members~\cite{DoyleLR92} (see also~\cite{Swanepoel18}).
If $M$ is a parallelogram, a linear map turns $B$ into $[-1,1]^2$, and $|S| \le 4$, because $v + [-1,1]^2$ is the union of four closed unit squares, each of which contains at most one point of $S$.
\end{proof}

\begin{proof}[Proof of \cref{thm:translates}]
By \cref{lem:translates,lem:translatesindep}, after a linear change of coordinates, which does not change the graphs, these graphs form a $3$-sweepable class with neighborhood independence at most $5$.
So \cref{thm:reuse} with $c = 3$, $a = 5$, $s = 27/20$, and $\varepsilon = 10^{-6}$ gives the second statement, as for \cref{thm:frac}, and the proof of \cref{thm:main}, which uses only these properties, gives the first.
\end{proof}

\section{A 1.68-Approximation Algorithm for Unit Square Graphs}\label{sec:squares}

This section proves \cref{thm:squares} and the lower bound $4/3$ on $\limsup \chi/\omega$ for unit square graphs.
Unit square graphs are $2$-sweepable with $\prec$ the lexicographic order on $(x,y)$ and the two closed unit squares $Q_1(v), Q_2(v)$ of \cref{lem:classes}.
For them the bound $2\chi$ of \cref{lem:slabs} is no better than the sweep, and we replace it by the following.

\begin{lemma}[overlapping slabs]\label{lem:stitch}
There is an algorithm that, given an integer $C \ge 2$ and a unit square graph $G$ on $n$ points with $\omega(G) \le C$ and at least one edge, colors $G$ with at most $\lceil 3\chi(G)/2 \rceil$ colors in $O(n \log n) + 2^{O(C \log^2 C)} n$ time.
\end{lemma}

\begin{figure}[t]
\centering
\includegraphics{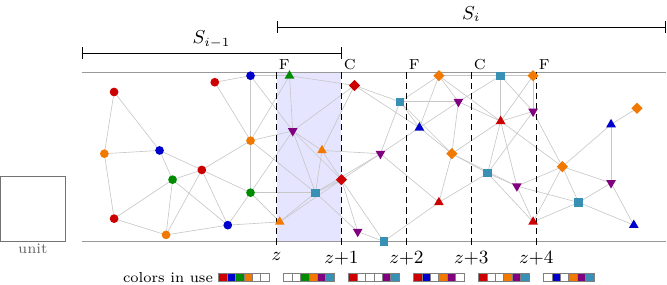}
\caption{\Cref{lem:stitch} with $k = 4$ and $h = 6$, near the left end $z$ of a slab $S_i$.
Colors show $d$, shapes the classes of $\sigma_i$, and circles the points outside $S_i$; the bars show the colors used in each strip.
The F-steps at $z$, $z+2$, and $z+4$ finish $2$, $1$, and $1$ classes, and to the right of $z + 4$ each class of $\sigma_i$ has a single color.}
\label{fig:stitch}
\end{figure}

We first outline the proof.
Overlapping vertical slabs of width $2\lceil \log_2 C \rceil + 2$ are colored optimally by \cref{lem:slabdp}; let $k \le \chi$ be the largest number of colors used, and use $h = \lceil 3k/2 \rceil$ colors.
The coloring of one slab is turned into that of the next at the cuts $z, z+1, \dots$, where $z$ is the left end of the next slab (\cref{fig:stitch}).
At $z$, $z+2$, \dots, whole color classes of the next coloring receive colors that are unused near the cut.
At $z+1$, $z+3$, \dots, the remaining points to the right of the cut are recolored by first-fit in the order $\prec$.
They are colored by the $\ell$ classes not yet repainted, so their clique number is at most $\ell$, and first-fit needs only $2\ell - 1$ colors, whatever the colors to the left of the cut.
The first cut finds $\lceil k/2 \rceil$ unused colors, and the number of unused colors then doubles at every other cut, so the next coloring is reached within distance $2\lceil \log_2 C \rceil$ of $z$, before the slab after it begins.

\begin{proof}
If two points that are consecutive in the order of $x$-coordinates differ by more than $1$ in $x$, the points on the two sides can be colored independently, so we assume that this does not happen.
We also assume that $G$ has an edge, since otherwise one color suffices.
Let $D = 2\lceil \log_2 C \rceil + 1$, let $x_0 = \min_{p \in P} x_p$, and for $i = 0, 1, \dots, \lfloor n/D \rfloor$ let the \emph{slab} $S_i$ be the set of points $p$ with $x_0 + iD \le x_p < x_0 + (i+1)D + 1$.
Consecutive slabs overlap in a strip of width $1$, every point lies in a slab, and every edge has both endpoints in some slab.

\emph{Coloring one slab.}
The points of $S_i$ lie in a vertical slab of width $D + 1$, so \cref{lem:slabdp} with $W = D + 1$ computes an optimal coloring $\sigma_i$ of $G[S_i]$.
Let $k = \max_i \chi(G[S_i])$, so that $2 \le k \le \chi(G)$ and, by \cref{lem:sweep}, $k \le 2C - 1$, and let $h = \lceil 3k/2 \rceil$, so that $k < h < 2k$.
All colors are taken from $[h]$.

\emph{Stitching.}
We define the coloring $d$ slab by slab, starting with $d = \sigma_0$ on $S_0$.
Before slab $i \ge 1$ is processed, with $z = x_0 + iD$, the following entry condition holds: $d$ is a proper coloring of the points with $x < z + 1$, it is undefined on the other points, and the points with $x \ge z - 1$ use only colors of a set $A \subseteq [h]$ of $k$ colors.
Let $B_1, \dots, B_k$ be the color classes of $\sigma_i$, some possibly empty.
We will assign some of them, called \emph{finished}, distinct colors $\varphi(B)$; the other classes and their points are \emph{unfinished}, and $f$ denotes the number of finished classes.
For $j = 0, 1, 2, \dots$ we perform a step at the \emph{cut} $z_j = z + j$, an \emph{F-step} if $j$ is even and a \emph{C-step} if $j$ is odd (\cref{fig:stitch}):
\begin{description}
\item[F-step.] Let $r = h - f - |A|$.
Give $\min\{r, k - f\}$ unfinished classes $B$ distinct colors $\varphi(B) \in [h] \setminus (A \cup \varphi(\text{finished classes}))$, mark them finished, and set $d(v) = \varphi(B)$ for every $v \in B$ with $x_v \ge z_j$.
If all classes are finished, stop.
\item[C-step.] Let $\ell = k - f$.
Replace $A$ by a set of $2\ell - 1$ colors of $[h] \setminus \varphi(\text{finished classes})$.
Recolor the unfinished points $u$ with $x_u \ge z_j$ by first-fit in the order $\prec$: $u$ gets the first color of $A$ that no earlier neighbor holds.
\end{description}
We show by induction on $j$ that after the step at $z_j$, with $z_{-1} = z - 1$ before the first step:
(a) $d$ is proper, and $A$ is disjoint from $\varphi(\text{finished classes})$;
(b) every point $v$ with $x_v \ge z_j$ at which $d$ is defined has $d(v) = \varphi(B)$ if $v$ lies in a finished class $B$, and $d(v) \in A$ otherwise;
(c) $d$ is defined on the points with $x < z + 1$ and on the finished points, after the first C-step on all of $S_i$, and nowhere else.
Before the first step this is the entry condition.
A step at $z_j$ changes only points $v \in S_i$ with $x_v \ge z_j$, and their neighbors $w$ have $x_w \ge z_j - 1 = z_{j-1}$, so (b) for the previous step applies to them; such a $w$ is outside $S_i$ only if $x_w < z$, which needs $j = 0$, or if $x_w \ge z + D + 1$, where $d$ is undefined by (c).

Consider an F-step and a recolored point $v \in B$.
Its neighbors hold colors of $A$, colors $\varphi(B')$ of classes finished earlier, or, if they were recolored in the same step, colors $\varphi(B')$ of other new classes $B' \neq B$, because $B$ is independent; none of these is $\varphi(B)$.
So (a) is kept, and (b) and (c) hold by construction.

Consider a C-step; then $j$ is odd and $z_j \ge z + 1$.
Let $u$ be an unfinished point with $x_u \ge z_j$; its neighbors have $x \ge z$, and those at which $d$ is defined lie in $S_i$.
Every earlier neighbor $w$ of $u$ has a color when $u$ is recolored: if $x_w < z_j$, by (c), since $x_w < z + 1$ or a C-step has already run; if $x_w \ge z_j$, because $w$ is finished or was recolored before $u$.
Every later neighbor has $x \ge x_u \ge z_j$, so it is finished or is recolored after $u$.
Finished neighbors hold colors outside the new $A$, by (b) for the previous step.
The unfinished points of $S_i$ induce a subgraph that $\sigma_i$ colors with $\ell$ colors, so its clique number is at most $\ell$; the earlier neighbors of $u$ lie in $Q_1(u) \cup Q_2(u)$, two cliques containing $u$, so at most $2(\ell - 1)$ of them are unfinished.
So first-fit finds a color in $A$, and every edge at a recolored point is proper: to a finished point because $A$ avoids $\varphi$, to an earlier unfinished point because $u$ avoided its color, and to a later point because that point avoided $d(u)$.
Again (a)--(c) follow.

It remains to see that the colors suffice and that the steps end in time.
At the first F-step $r = h - k = \lceil k/2 \rceil$, so it finishes $\lceil k/2 \rceil$ classes and leaves $\ell_1 = \lfloor k/2 \rfloor \ge 1$.
Before every later F-step, $|A| = 2\ell - 1$ with $\ell = k - f$, so $r = E(f) := h - 2k + f + 1$, and the C-step before it needs $E(f) \ge 0$.
Now $E(\lceil k/2 \rceil) = 2h - 3k + 1 \ge 1$, and $E$ is increasing in $f$, so every C-step finds its colors and every later F-step has $r \ge 1$.
If an F-step finishes $r = E(f) < \ell$ classes, the next one finds $E(f + r) = 2r$ unused colors.
So after the first F-step and $J$ further F-steps all classes are finished as soon as $2^J - 1 \ge \ell_1$.
Since $\ell_1 = \lfloor k/2 \rfloor \le C - 1$, $J = \lceil \log_2 C \rceil$ suffices, and the last step is the F-step at a cut $z_j \le z + 2J = z + D - 1$.
Then every class is finished, so by (b) and (c) the points with $x \ge z + D - 1$ at which $d$ is defined use only the $k$ colors $\varphi(B_1), \dots, \varphi(B_k)$, and $d$ is a proper coloring of the points with $x < z + D + 1$ and undefined elsewhere.
This is the entry condition for slab $i + 1$, with $z + D$ in place of $z$.
After the last slab, $d$ is a proper coloring of $G$ with at most $h = \lceil 3k/2 \rceil \le \lceil 3\chi(G)/2 \rceil$ colors.
Each point is recolored $O(D)$ times, each time checking its fewer than $25C$ neighbors, which a sweep lists for all points in $O(n \log n + Cn)$ time, so the stitching takes $O(n \log n + nC \log C)$ time.
Every point lies in at most two slabs and $D = O(\log C)$, so by \cref{lem:slabdp} the slabs are colored in $O(n \log n) + 2^{O(C \log^2 C)} n$ time in total.
\end{proof}

The proof uses only that each $\sigma_i$ is a proper coloring with $k$ colors; optimality is needed only for $k \le \chi(G)$.
So every unit square graph satisfies $\chi(G) \le \lceil \tfrac32 \max_i \chi(G[S_i]) \rceil$ for slabs $S_i$ of width $2\lceil \log_2 \omega(G) \rceil + 2$, whatever $\omega(G)$ is.
The proof also works for every $c$-sweepable class whose order $\prec$ satisfies $u \prec v$ whenever $x_u < x_v$, with $h = \lceil ((2c-1)k - c + 2)/c \rceil$ colors and sets $A$ of $c(\ell - 1) + 1$ colors, and then uses at most $(2 - \tfrac{1}{c+1})\chi$ colors; for unit disk graphs this is $\tfrac74\chi$, which we do not need.
Since $\lceil 3\chi/2 \rceil \le \tfrac53\chi$ for $\chi \ge 2$ and $\tfrac53 < 1 + \ln 2 < 1.6932$, the proof of \cref{lem:absolute} with $c = 2$, $\varepsilon = 1/1000$, and \cref{lem:stitch} in place of \cref{lem:slabs} colors every unit square graph with fewer than $1.6942\,\chi$ colors, without reuse.

\subparagraph*{Reuse.}
Unit square graphs have neighborhood independence at most $4$, since $N_G(v)$ lies in $v + [-1,1]^2$, which is the union of four closed unit squares, each inducing a clique (\cref{fig:shapes}(d)).
For $c = 2$, $a = 4$, and $s = 4/5$ we get $2e^{-s} \ge \Phi_4(s)$ and $\kappa_{2,4}(s) < 1.679599$, within $10^{-4}$ of the minimum over $s$, while $1 + \ln 2 > 1.6931$.

\begin{proof}[Proof of \cref{thm:squares}]
\Cref{thm:reuse} with $c = 2$, $a = 4$, $s = 4/5$, and $\varepsilon = 10^{-6}$ gives the second statement.
For the first, the proof of \cref{thm:main} with these parameters and \cref{lem:stitch} in place of \cref{lem:slabs} uses at most $(\kappa_{2,4}(4/5) + 10^{-6})\chi < 1.6796\,\chi$ colors if $F \ge C$.
Otherwise \cref{lem:stitch} uses at most $\lceil 3\chi/2 \rceil \le \tfrac53\chi$ colors if $\chi \ge 2$, one color suffices if $\chi = 1$, and $\tfrac53 < \kappa_{2,4}(4/5)$.
\end{proof}

\begin{figure}[t]
\centering
\includegraphics{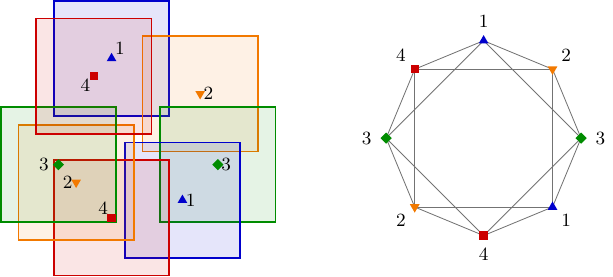}
\caption{The graph $C_8^2$, with $i$ adjacent to $i \pm 1$ and $i \pm 2$ modulo $8$, as eight unit squares (left) and on a circle (right), with a $4$-coloring.
Since $\omega = 3$ and $\alpha = 2$, we have $\chi \ge 8/2 = 4$.}
\label{fig:c8}
\end{figure}

\subparagraph*{A lower bound on $\chi/\omega$.}
The published lower bound $5/4$ on $\limsup \chi/\omega$ for unit square graphs is obtained from the $5$-cycle $C_5$ by replacing every square with $t$ copies of itself~\cite{DumitrescuJ12,CaoduroS24}.
The square of the $8$-cycle, $C_8^2$, with vertices $0, \dots, 7$ and $i$ adjacent to $i \pm 1$ and $i \pm 2$ modulo $8$, is the member after $C_5$ of the cycle-power family used for disks~\cite{MalesinskaPW98}, and it is also a unit square graph (\cref{fig:c8}).
It has $\omega = 3$ and $\alpha = 2$, so taking $t$ copies of every square gives unit square graphs with $\omega = 3t$ and $\alpha = 2$, and hence $\chi \ge 8t/2 = 4t$.

\bibliography{refs}

\section*{AI Statement}
This paper was prepared with substantial help from the AI assistants Claude (Anthropic) and ChatGPT (OpenAI).
They took part in developing the algorithms and proofs, checking the references, and writing the text.
The authors checked all results and are responsible for the content.

\end{document}